\documentclass[aps,twocolumn,superscriptaddress]{revtex4-1}
\usepackage{graphics} 
\usepackage{epsfig} 
\usepackage{amsmath}
\usepackage[T1]{fontenc}
\usepackage[latin9]{inputenc}
\usepackage{color}

\begin{document}
\title{Quasilocalized charge approximation approach for the nonlinear structures in strongly coupled Yukawa systems}
\author{Prince Kumar}
\email{prince.kumar@ipr.res.in}
\affiliation{Institute for Plasma Research, Bhat, Gandhinagar, India, 382428}
\affiliation{Homi Bhabha National Institute, Training School Complex, Anushaktinagar, Mumbai 400094, India}
\author{Devendra Sharma}
\email{devendra@ipr.res.in}
\affiliation{Institute for Plasma Research, Bhat, Gandhinagar, India, 382428}
\affiliation{Homi Bhabha National Institute, Training School Complex, Anushaktinagar, Mumbai 400094, India}
\date{\today}

\begin{abstract}
Strongly coupled systems occupying the transitional range between the Wigner 
crystal and fluid phases are most dynamic constituents of the nature. 
Highly localized but strongly interacting elements in this phase
posses enough thermal energy to trigger the transition between 
a variety of short to long range order phases. 
Nonlinear excitations are often the carriers of proliferating structural 
modifications in the strongly coupled Yukawa systems.
Well represented by a laboratory dusty plasma, these systems show explicit 
propagation of nonlinear shocks and solitary structures both in experiments 
and in first principle simulations. 
The shorter scale length contributions remain absent at strong screening 
in present approximate models which nevertheless
prescribe nonlinear solitary solutions that consequently lose their 
coherence in a numerical evolution of the system under a special 
implementation of the quasi-localized charge approximation formulation. 
The stable coherent structures self-consistently emerge following an 
initial transient in the numerical evolution which adapts QLCA approach to 
spatiotemporal domain for accessing the nonlinear excitations 
in the strong screening limit.
The present $\kappa\sim 1$ limit of the existing Yukawa fluid models to 
show agreement with the experiment and MD simulations has therefore been 
overcome and the coherent nonlinear excitaitons have become characterizable
up to $\kappa\sim 2.7$, before they becoming computationally challenging in 
present implementation. 

\end{abstract}



\maketitle
\section{Introduction}
Physical systems exhibiting a strongly coupled 
phase in the limit of potential energy of interaction exceeding
the kinetic or random energy of particles are very effectively modelled by
a highly charged minority dust species immersed in an electron-ion plasma. 
The deterministic configurational correlation between these highly charged 
species often dominates their response over the kinetic randomness. 
This most accessible example of a strongly coupled systems allows 
quantitative description of the possible solid and gaseous phases by 
only two parameters, namely, the coupling parameter 
$\Gamma=Z^{2}e^{2}/ak_{B}T$ and the screening parameter 
$\kappa=a/\lambda_{D}$, where $\lambda_{D}$ is the plasma Debye length, 
$a$ is inter-dust separation (Wigner-Seitz radius), $Z$ is dust charge 
multiplicity of the electronic charge $e$, $T$ is dust kinetic temperature 
and $k_{\rm B}$ is the Boltzmann constant.
A vast range of strongly coupled systems modeled by this include
dense astrophysical systems \citep{peng2007}, warm dense matter, 
trapped ions \citep{dantan2010}, ultracold neutral plasma \citep{killian2007}, 
etc. 

An intermediate state between the so called Wigner crystal \citep{Wang_2001} 
($\Gamma > \Gamma_{\rm melting}$) phase and the gaseous phase ($\Gamma \sim 0$)
remains the most challenging one. This is evident from the fact that a 
dynamic mean-free version of the random phase (gaseous) approximation 
\citep{hou2004,hou2009,kaw1998,kaw2001collective}, 
as well as a solid phase approach acknowledging particle correlation 
function $g(r)$ \citep{Golden,donko2008dynamical} 
have been applied to this intermediate phase with limited success.
In accounting for $g(r)$, a 
considerable localized stay of dust particles in their fragile 
(piecewise stable)
macroscopic potential landscape is acknowledged just below 
$\Gamma_{\rm melting}$ closely resembling the phonon formulation applicable 
well above $\Gamma_{\rm melting}$. 
This Quasi-localized Charge Approximation (QLCA) 
is thus applied with a considerable success not only in linear perturbative 
limit 
\citep{Golden,Kalman_2005} but 
finite prospects are also explored of its non-perturbative \citep{golden2009collective} applicability 
to the
nonlinear excitations 
\citep{lee2011nonlinear}.



In a laboratory setup, dusty plasmas exhibit both molten and crystelline
phases of a strongly coupled system \citep{morfill2009complex}.  The linear excitations of the dust, namely the
the longitudinal dust acoustic waves, and transverse shear modes are well 
explored by various theoretical models, 
such as 
the Generalized Hydrodynamics (GH)\citep{kaw1998,kaw2001collective}, thermodynamic approach\citep{khrapak2015fluid}, $T^{\rm (eff)}$ model\citep{yaroshenko2010}
and Quasi-localized Charge Approximation (QLCA) \citep{Golden,khrapak2016long,kalman2000collective}.
In the weak screening regime $\kappa$ (= a/$\lambda_{D}$) $\leq$ 1 (where $a$ 
being the lattice constant and $\lambda_{D}$ being the debye length), 
the GH approach described not only of the longudinal modes but also reproduced the gap at long wavelengths, or the $k$ gap \citep{kaw1998,kaw2001collective}, of the transverse (shear) acoustic mode dispersion.
The excitation in the strong screening limit ($\kappa \gg 1$) accessed by GH 
model however depart from the MD simulation dispersion \citep{ohta2000wave} 
as the original One Component Plasma (OCP) version of the GH model 
requires essential screening specific corrections both in the equation of 
state and the excess energy $u(\Gamma)$ which enter its phenomenological 
dispersion relation \citep{kaw1998,kaw2001collective}.
On the other 
hand, linear dispersions in the strong coupling limit are successfully 
recovered under the QLCA formulation and the resulting dispersion 
\citep{Golden,khrapak2016long,donko2008dynamical} 
remains in agreement with experiments \citep{nunomura2005wave} and MD 
simulations 
\citep{ohta2000wave,khrapak2016long,donko2008dynamical,kalman2000collective}.
The QLCA formulation, by accounting for a qasi-localized dust structure,
omits dust diffusion to work in the infinite relaxation time 
limit ($\tau_{r} \rightarrow \infty$) of the phenomenological dispersion 
$\omega^{2}=c^{2}k^{2}-\tau_{r}^{-2}$ \citep{yang2017emergence}, hence excludes the $k$-gap of the 
transverse dispersion and access $k\rightarrow 0$ excitations.
The nonlinear longitudinal excitations, which are subject of the present 
paper, are treated for the first time under QLCA formulation which besides
being in agreement with GH solutions in weak screening limit 
\cite{sharma2014head,boruah2016observation}, 
allows access to the presently unexplored strong screening limit of the 
nonlinear excitations.

Despite its strength in treating the strong screening regime, the application 
of QLCA formulation to nonlinear excitations is limited by
its intrinsically spectral form largely suitable to linear excitations. 
This barrier is overcome in the present treatment by adopting the 
recently developed excluded volume approximation \citep{khrapak2016long} 
and its numerical implementation for evaluating the QLCA dynamical matrix 
in the spatiotemporal domain for ready applicability to nonlinear 
perturbations and exmining their stability with respect to temporal evolution. 
We have first shown that an analytical 
approximation of the QLCA dynamic matrix 
$D_{L}$\citep{Golden,rosenberg1997}
in terms of excess energy (the OCP implementation), limited to weak screening 
limit, reproduces results available from the GH prescription. 
Applying QLCA formulation in its full capacity to the strong screening limit, 
we subsequently recover a strong departure of the nonlinear excitations 
from their weak screening counterparts. The 
nonlinear structures in this limit are shown to be characteristically 
distinct when compared to the OCP implementation of the formulation 
when the latter is nevertheless used to produce nonlinear solutions with 
relatively larger $\kappa$.
The access to larger frequency, or shorter wavelength limit, where the OCP 
based descriptions, including the original GH dispersion as well as its QLCA
counterpart, show strong limitation is now available by means of the presented 
excluded volume approximation of the QLCA dynamical matrix. The corresponding 
QLCA linear dispersion is shown to closely agree with the results of the MD 
simulation in this regime.
In particular, the $\kappa\sim 1$ limit for the existing
Yukawa fluid models to show agreement with the experiment 
and simulations has been overcome by the present QLCA prescription and 
agreement is now recoverable up to $\kappa\sim 2.7$. 

The article is organized as follows. In Sec.~\ref{qlca}, the linear QLCA 
formulation is outlined beginning from a more general rotating frame version 
of it analyzed recently \citep{kumar2021collective}. 
The nonlinear approach to excitations with 
localization accounted for by QLCA formulation within the KdV framework 
as well as in a general pseudospectral nuemrical framework are presented in 
Sec.~\ref{KdV} and Se.~\ref{simulation}, respectively, alongwith a 
description of 
isothermal dust compressibility in Sec.~\ref{compressibility}. 
Impact of the localization on linear and nonlinear analytical solutions of 
the model is analyzed on KdV-prescribed and more general form of coherent 
perturbations in Sec.~\ref{simulation} and Sec.~\ref{results}, 
respectively.
Summary and conclusions are presented in Sec.~\ref{conclusion}
\section{Nonlinear QLCA theory for a strongly coupled Yukawa system \label{qlca}}
The QLCA theory has been successfully applied to study collective excitation of the liquid phase strongly coupled systems in a rotating \citep{kumar2021collective} as well as non-rotating frame\citep{Golden}. The microscopic equation of motion of the dust particle in a rotating frame,  for the component $r_{i\mu}$ aligned to the direction $\mu~(=x,y)$,
 \begin{eqnarray}
\begin{split}
m_{\rm d}\frac{\partial^{2}{r}_{i\mu}}{\partial t^2}=
	\sum_{j} {K}_{ij\mu\nu}r_{j\nu}
	-2m_{\rm d}\left[{\bf \Omega} \times \frac{\partial {\bf r}_{i}}{\partial t}\right]_{\mu}\\
	-m_{\rm d}[{\bf \Omega}\times({\bf \Omega}\times {\bf r}_{i})]_{\mu} 
	-\frac{\partial V}{\partial r_{\mu}}=0,
\end{split}
	\label{particle-eq_rotating}
\end{eqnarray} 
where the second and third terms in the right-hand side are the Coriolis force and centrifugal force, respectively. The quantity V is the dust confinement potential whose  gradient balances the corresponding component of the centrifugal force in the typical equilibrium condition \citep{kumar2021collective}.
This equation can be reduced, in a non-rotating frame ($\Omega\rightarrow 0$), to a form 
given as \citep{Golden}, 
 \begin{eqnarray}
\begin{split}
m_{\rm d}\frac{\partial^{2}{r}_{i\mu}}{\partial t^2}=
	\sum_{j} {K}_{ij\mu\nu}r_{j\nu},
\end{split}
	\label{particle-eq}
\end{eqnarray}
where the non-retarded limit of ${K}_{ij\mu\nu}$ defines the potential 
energy of the strongly coupled Yukawa fluid. The particles interact with each 
other through a shielded potential, namely, the Yukawa potential which is 
provided by the uncorrelated background plasma, given as,
\begin{eqnarray}
	\phi(|{\bf r}_i-{\bf r}_j|)=e^{-\kappa_{b}|{\bf r}_{i}-{\bf r}_{j}|}\frac{Z^{2}e^{2}}{|{\bf r}_{i}-{\bf r}_{j}|},
\end{eqnarray}
where the screening parameter $\kappa_{b}$ is defined by the uncorrelated background plasma pressure and 
self consistent electric field 
\textbf{E} = $\nabla \phi(|{\bf r}_i-{\bf r}_j|)$ between the negatively 
charged dust particles can be derived. 

The well known linear QLCA results are recovered from equation for 
the linear perturbation eigenmodes $\xi_{{\bf k}\nu}(\omega)$ of $r_{i\mu}$
and an ensemble averaging of the collective coordinates
\cite{Golden,Kalman_2005},
\begin{eqnarray}
	\nonumber
	[\omega^{2}\delta_{\mu\nu}-C_{\mu\nu}({\bf k},\omega)]
	\xi_{{\bf k}\nu}(\omega)
	=0,
\label{linear-equation}
\end{eqnarray}
where $C_{\mu\nu}$ contains the mean field and local field effects, 
produced by the random motion and the dust-dust correlation, 
respectively,
\begin{eqnarray}
	C_{\mu\nu}({\bf k},\omega)
	=\omega_{\rm pd}^{2}
	\left[\frac{k_{\mu}k_{\nu}}{k^{2}+\kappa_{b}^{2}}
	+\cal{D}_{\mu\nu}({\bf k},\omega)\right],
	\label{dispersion-qlca}
\end{eqnarray}
with $\omega_{pd}$ being the dust acoustic frequency. 
The central quantity in the QLCA, 
the dynamical matrix in three dimensions, given as,
\begin{eqnarray}
	D_{L,T}({\bf k})=\omega_{\rm pd}^{2}\int_{0}^{\infty}dr
	\frac{e^{-\kappa r}}{r}
	[g(r)-1]{\cal K}_{L,T}(kr,\kappa r), 
	\label{D-long-wave-length}
\end{eqnarray}
includes dust-dust 
correlation effects, as this is a functional of the equilibrium pair 
correlation function (PCF). 
Clearly, in the absence of any correlations $g(r)\rightarrow 1$ at all values 
of $r$, the linear QLCA approach begins to provide the mean-field, 
or random-phase 
approximation results ($\cal{D}_{\mu\nu}$ = 0). 
Since the desired strong coupling effect enter the formulation predominantly 
by means of the functional $D_{L,T}$, the nonlinear treatment presented here 
exploits its adoptation to the spatiotemporal domain while retaining the
contributions from the coupling of the individual modes, exclusively 
achievable under a nonlinear pseudospectral framework\citep{fornberg1998practical} .
\subsection{Strong coupling in explicit QLCA approach}
There remain two options for 
including the strong coupling effects applicable to two different states of 
the dust medium, namely, the strong coupling without localization ($g(r)=1$) 
and the stronger coupling with finite localization ($g(r)\neq 1$). 
While the first is achievable by an ad hoc inclusion of strong coupling 
effects in the pure fluid approach without invoking the QLCA framework since 
there is no localization, however the second essentially requires QLCA 
framework as $g(r)\neq 1$ and it is impossible to reduce QLCA to the fluid 
theory for finite localization limit without reasonable approximations. 
This puts a natural limit on applicability of 
the first kind of approach to strongly coupled systems, since when coupling is 
sufficiently strong finite localization must emerge and the second kind of 
approach begins to be applicable \citep{Kalman_2005,donko2003molecular}. 
We have summarized the (linear) dispersions from treatments
belonging to the first option in Appendix~\ref{A} their relation will be 
discussed with the nonlinear results 
obtained by us using the explicit QLCA approach ($g(r)\neq 1$) 
which show distinction with the existing nonlinear results. 
%

\begin{figure}[hbt]
 \centering
 \includegraphics[width=80mm]{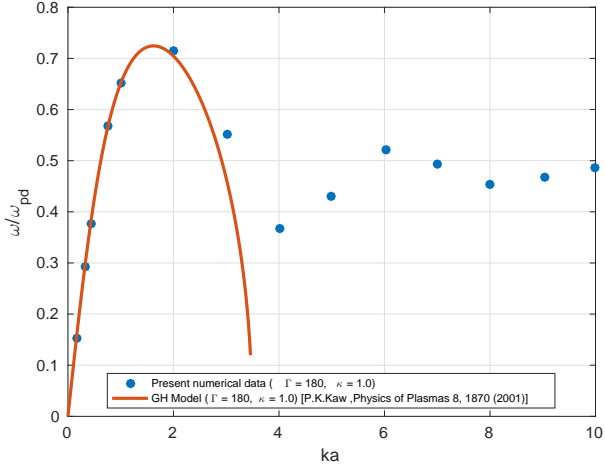}
  \caption{The red curve crossespond to the dispersion relation obtained from  the GH model \citep{kaw2001collective} and the blue dots are
plotted using QLCA based model, for same parameter values, $\Gamma$ = 180 and $\kappa$ = 1.0. }
 \label{GH_Model_QLCA}
\end{figure}

\begin{figure}[hbt]
 \centering
 \includegraphics[width=90mm]{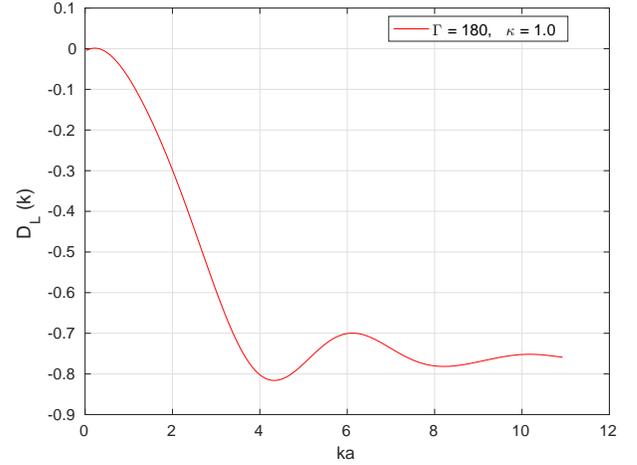}
  \caption{ The plot of $D_{L}$-matrix using  equation \eqref{eq:Full_D} which is obtained from the MD data and dispersion relation given in \citep{khrapak2016long} with $\kappa$ = 1. }
 \label{D_matrix}
\end{figure}

In the limit where $D_{L}$ is a function of excluded volume parameter
R($\Gamma$,$\kappa$) 
\citep{khrapak2016long}, it is derived by choosing a simple but reasonable 
valid approximation on $g(r)$, i.e.,  $g(r) = 1$ for $ r > R $ and $g(r) = 0$ 
otherwise, or in a long-wavelength limit, as,
\begin{eqnarray}
 \nonumber
	D_{L}= -\omega^{2}_{0}(k)+\omega_{\rm pd}^{2}e^{-k R}\left[
	(1+k R)\left(\frac{1}{3}-\frac{2\cos{kR}}{k^2R^2}
	\right.\right.\\
	\left.\left.
	+\frac{2\sin{kR}}{k^3R^3}\right)
	-\frac{\kappa^{2}}{\kappa^{2}+k^{2}}
	\left(\cos{kR}+\frac{\kappa}{k}\sin{kR}\right)\right] \nonumber
	\\
\label{eq:D_matrix}
\end{eqnarray}
%
%
\begin{figure}[hbt]
 \centering
 \includegraphics[width=80mm]{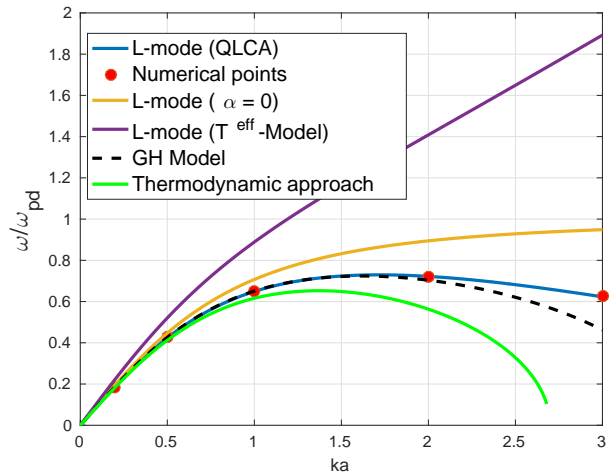}
  \caption{ Analytical (blue lines) and numerical (red dot points) longitudinal wave dispersion relation for the charged dust fluid with values $\Gamma$ = 185, $\kappa$ = 1.0, and excluded volume $R(\kappa)$ = 1.1. The dispersion relation represented by the yellow and purple lines are obtained from a weakly coupled limit ($D_{L}$ = 0) and from a  $T^{\rm (eff)}$ model, respectively. The green curve and dash line correspond to the dispersion relation  calculated using the thermodynamic approach and GH model, respectively, for the parameters $\kappa$ = 1 and $\Gamma$ = 207.}
 \label{dispersion-comparison}
\end{figure}
The effective parameter for the excluded volume, $R(\Gamma,\kappa)$, is 
function of 
the system state variables, $\kappa$ and $\Gamma$, evaluated using the 
expression of a system correlation energy. A simple explicit expression for 
this excluded volume parameter,  $R\approx 1+ \frac{\kappa}{10}$, has been 
calculated for a weak screening regime \citep{khrapak2017practical}.
In Fig.~\ref{dispersion-comparison} we have compared the 
strong-coupling limit linear QLCA dispersion, Eq.
(\ref{dispersion-qlca}), with the dispersion in Eq.
(\ref{dispersion-rpa}), Eq. (\ref{dispersion-thermo}) and GH model 
\citep{kaw2001collective} which admit 
the strong coupling effects using $T^{\rm (eff)}$, thermodynamic functions and  viscoelastic coefficients, respectively, but no explicit localization.
The QLCA dispersion relation in Eq. (\ref{dispersion-qlca}) as computed
under excluded volume approximation for the case using $\Gamma$ $\approx$ 180, 
$\kappa$ = 1 and R = 1.1 as in Ref.~\citep{khrapak2016long} is plotted 
in Fig.~\ref{dispersion-comparison} using blue line which is in 
agreement with Ref.~\citep{khrapak2016long}.
The red dots superimposed on the plot are numerically computed dispersion
relation by solving, using pseudo-spectral method, the full nonlinear set of 
QLCA equations under the same approximation showing excellent agreement with 
linear model at small amplitudes.
While the yellow curve saturating to dust acoustic frequency $\omega_{\rm pd}$ 
represents the pure random-phase (fluid-limit) dispersion relation, a
contrasting behavior with respect to QLCA dispersion is shown by the 
dispersion plotted with purple line which is the strong coupling dispersion 
(\ref{dispersion-rpa}) not admitting localization. While the green curve, obtained from the Eq.(\ref{dispersion-thermo})  
 shows correspondence with the QLCA model (blue curve/red dots) at very long wavelength regime ($ ka< $ 1.0), it shows continuous deviation from the QLCA dispersion at relatively higher mode number (k $\ge$ 1.0), as can be seen form the Fig.~\ref{dispersion-comparison}. 
Note that both the analytical dispersion and the numerical 
solutions (red dots) obtained from full nonlinear QLCA model implemented by 
us (red dots in Fig.~\ref{dispersion-comparison}) show agreement also with the
Molecular Dynamical(MD) simulations \citep{donko2008dynamical,khrapak2016long}
as well as the experimental studies \citep{bandyopadhyay2007experimental}.
While the GH model provides good agreement with the QLCA based model at long wavelengths and weak screening limit, it requires presently unavailable corrections for agreement with shorter wavelength and strong screening limit of the dispersion relation recovered in MD simulations \cite{kaw2001collective}. 
Access to this limit is additionally shown to be possible by a numerical 
implementation of $D_{L}$ under the QLCA framework where a more sophisticated
form of $g(r)$, derived from MD simulation data, can be used rather than the 
excluded volume approximation. Results from this implementation are shown 
to make the comparison with GH model possible up to larger $k$ values in 
Fig.~\ref{GH_Model_QLCA} where
%
the radial pair correlation function $g(r)$ used in our pseudospectral 
computations is obtained from the MD 
simulation data adopted from Ref.~\citep{khrapak2016long}. This yields the 
expression for $D_{L}$-matrix, 
\begin{eqnarray}
 \nonumber
D_{L}(k)	= \omega^2_{pd} [(-0.568 + 0.3149 \cos(0.431 k) + \\
 0.056\sin(0.431 k) \nonumber
 +0.187\cos(0.862k) \\ \nonumber  + 0.058\sin(0.862k) + 0.0511 \cos(1.293k) \\ \nonumber
  + 0.044\sin(1.293k) -0.00016\cos(1.724k)  \\  \nonumber
  - 0.025\sin(1.72k)+ 0.0079\cos(2.155k) \\
   - 0.0090\sin(2.155k))] 
\label{eq:Full_D}
\end{eqnarray}
which is also plotted in 
Fig.~\ref{D_matrix} with state parameters $\Gamma$ = 180 and $\kappa$ = 1.0.  

While, as shown in Fig.~\ref{GH_Model_QLCA}, the GH Model and QLCA predict an 
identical dispersion relation in a long-wavelength ($ka \le$ 2) limit, in a 
relatively short wavelength limit the present QLCA based numerical simulations 
successfully reproduce the oscillatory behavior which is characteristically 
observed in various theoretical studies 
\citep{khrapak2016long,donko2008dynamical}, MD simulations 
\citep{mokshin2022self} and experiments \citep{zhdanov2003,nunomura2005wave}. 
It is therefore clear that the result from the presently available OCP version 
of the GH model are not a reliable predictor for the short wavelength 
(high $k$) excitations.  
%
%
%
%
The corresponding $D_{L}(k)$ plotted in Fig.~\ref{D_matrix} 
shows that in the long wavelength regime (small $k$) the $D_{L}(k)$ has 
a variation which corresponds to the $g(r>R) \simeq$ 1 
as approximated by the excluded volume prescription \cite{khrapak2016long}. 
Under this approximation the large $r$ contribution to the 
wave dynamics dominates over the local interactions. 
At sufficiently shorter wavelengths (large $ka$),
the strength of $|D_{L}|$ increases considerably, meaning that the structural 
effect contribution (i.e., $g(r)$ featuring multiple peaks in 
Ref.~\cite{khrapak2016long}) to the wave dynamics 
increase significantly, an effect uniquely accounted for by the QLCA 
formulation. 

%
 %
 %
 \begin{figure}[hbt]
 \centering
 \includegraphics[width=80mm]{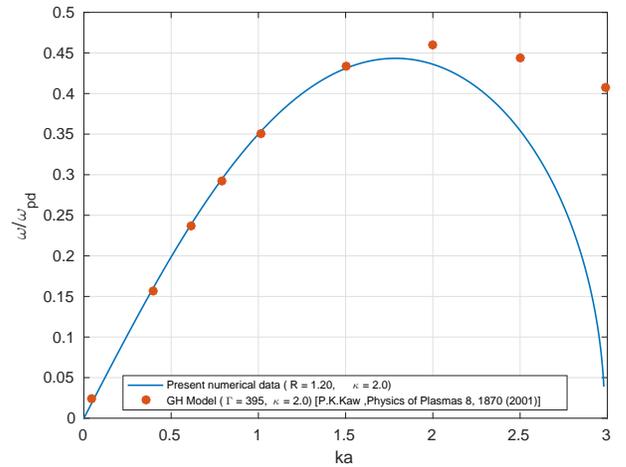}
  \caption{The blue curve crossespond to the dispersion relation obtained from  the GH model \citep{kaw2001collective}, $\Gamma$ = 395 and $\kappa$ = 2.0 and the blue dots are
plotted using QLCA based model equation (\ref{dispersion-qlca}) for value, R=1.20 and $\kappa$ = 2.0.}
 \label{QLCA_Kaw}
\end{figure}
For the shorter wavelength excitations, however, the discrepancy between 
the QLCA simulation based predictions (blue dots) and those of the GH model 
(solid curve) becomes significant. 
Remarkably, the excluded volume approximation, which uses a simplified step
function like profile for the $g(r)$, still produces reasonably good correction 
to the linear dispersion, as in Fig.~\ref{dispersion-comparison}, and 
prescribes frequency values that recover from their steep drop 
at larger $k$, to instead saturate to the Einstein frequency \cite{wong2017einstein,donko2008dynamical}.
This also remains the reason for a 
relatively better agreement between even approximate linear QLCA results 
and the MD simulation data.
The comparison between the dispersions from GH and QLCA (excluded volume) 
for an additional case with higher screening parameter $\kappa=2.0$ is done 
in Fig.~\ref{QLCA_Kaw}. In both cases the QLCA results remain in agreement 
also with MD simulations \citep{ohta2000wave} (not reproduced here) up to 
$ka\sim 3$.
\subsection{Nonlinear excitations of a strongly coupled localized phase dusty 
plasma\label{KdV}}
The central idea of localization involves the macroscopic variables 
obtained from the ensemble averaged particle equations. For a nonlinear 
approach it is however required that the averages are done over spatiotemporal
functions rather than their Fourier transformations. 
We therefore let the fluid conservation equations represent the 
evolution of these ensemble averages. We however acknowledge the presence of
a spatial ordering of the dust sites by allowing the dynamical matrix 
$D_{L,T}$ (determining mechanical response of the system) to be computed 
via the grain-grain correlation energy which changes based on the strain in 
spatial ordering, in addition to the routine (random phase) response arising 
from the associated background plasma compression.

This ensemble averaged (macroscopic) momentum equation has same form as 
Eq.~(\ref{momentum-balance-rpa}),
 \begin{eqnarray}
  \begin{split}
	 \frac{\partial{u}_{{\rm d}x} }{\partial t} + ({\bf u}_{{\rm d}}
	 \cdot\nabla) {u}_{{\rm d}x} 
	 = \frac{q_{\rm d}}{m_{\rm d}}{E}_{x} - \frac{1}{m_{d} n_{d}}
	  \frac{\partial P_{di}}{\partial x},
	  \label{momentum-balance-qlca}
\end{split}
\end{eqnarray}
where the dust kinetic energy is neglected because of being 
a few orders smaller than the representative strong coupling term  
$\frac{\partial P_{di}}{\partial x}$.
Similarly, the macroscopic particle continuity equation obtained by ensemble 
averaging over the dust sites is, 
\begin{equation}
	\frac{\partial n_{dx}}{\partial t} + \frac{\partial}{\partial x}
	(n_{dx} u_{dx}) = 0.
\label{continuity}
\end{equation}
Separate from the electric field $E_{x}$ produced by plasma species, the 
second term in the RHS of 
Eq. (\ref{momentum-balance-qlca}) accounts for the electrostatic field 
produced by the collective shift of the dust particles from their localized
positions which must be predominantly restored by the dust structure, rather 
than purely by $E_{x}$.
The term is therefore representable as a product of the density 
gradient produced and a force per unit density gradient. The latter is a rather 
sophisticated, localization-based, isothermal dust compressibility, treated 
further, both analytically and numerically, in Sec.~\ref{compressibility}.
It can be noted that, by construction, 
this new contribution must vanish if (i) there is no structured
background dust distribution (i.e., if $g(r)\rightarrow 1$) and/or 
(ii) if these is no dust density perturbation with respect to this uniform
structured dust background (i.e., if $\partial n_{d}/\partial x=0$). 
The Eq.~(\ref{momentum-balance-qlca}) can therefore be written as, 
%
\begin{equation}
\frac{\partial{u}_{dx} }{\partial t} +
{u}_{dx}\frac{\partial{u}_{dx} }{\partial x} = 
	-\frac{q_{d}}{m_{d}}\frac{\partial\phi}{\partial x}
	-\alpha \frac{1}{n_{d}}\frac{\partial n_{d}}{\partial x}. 
	\label{momentun-balance-qlca-2}
\end{equation}
where $\alpha_{c}=\frac{1}{m_{d}} \frac{\partial P_{di}}{\partial n_{d}}$ 
is the isothermal compressibility of the dust.
Up on normalization, Eq.~(\ref{momentun-balance-qlca-2}) takes the form,
\begin{equation}
\frac{\partial{u}_{dx} }{\partial t} +
{u}_{dx}\frac{\partial{u}_{dx} }{\partial x} = 
	\mu\frac{\partial\phi}{\partial x}
	-\tilde{\alpha} \frac{1}{n_{d}}\frac{\partial n_{d}}{\partial x},
	\label{momentun-balance-qlca-3}
\end{equation}
and the normalized Poisson equation becomes,
\begin{equation}
  \frac{\partial^2\phi}{\partial x^2} = \frac{1}{\mu}\left[n_{d}+n_{e}-n_{i}\right],
\label{poisson-equation}
\end{equation}
are now taken to be normalized. 
where
we have $\mu=\frac{Z_{d} T_{i}}{m_{d} a^{2}\omega^{2}_{pd}}$,
$\tilde{\alpha}=\alpha/\omega_{\rm pd}^{2}a^{2}$,
$\delta_{e}=n_{e0}/Z_{d}n_{d0}$, 
$\delta_{i}=n_{i0}/Z_{d}n_{d0}$,
$n_{e}=\delta_{e}(\sigma_{i}\phi)$,
$n_{i}=\delta_{i}(-\phi)$ and $\sigma_{i}$ = $T_{i}/T_{e}$,
while the equilibrium dust density $n_{d0}$, the ratio $T_{i}/e$, inverse 
dust acoustic frequency $\omega_{\rm pd}^{-1}$, and mean dust separation $a$ 
are used as normalizations for the density, potential, time and length, 
respectively.
Eq.~(\ref{momentun-balance-qlca-3})-(\ref{poisson-equation}) along with the 
continuity Eq.~(\ref{continuity}) constitute a 
nonlinear model. 
Subject to an accurate representation of $\alpha$ in temrs of $D_{L}$,
this can either be solved numerically, as 
done by means of the pseudospectral simulation procedure in the present work,
or in a rather approximate (and routine) analytical procedure of the reductive 
perturbation theory \citep{davidson2012methods} (duly done in Sec.~\ref{rpt}, obtaining the associated KdV equation and its solutions).

The methodology followed here is to first obtain the solutions of the 
nonlinear KdV equation and use them as initial profiles in our spatiotemporal 
pseudospectral numerical evolution which covers additional, 
previously uncovered, parameter regime. More general initial profiles 
are numerically evolved after this analysis.
%
%
\subsubsection{Derivation of the KdV equation for a strongly coupled fluid \label{rpt}}
In order to obtain the KdV equation \citep{davidson2012methods} for this system, we first introduce slow 
variable $\zeta$ and $\tau$, given by,
\begin{equation}
\zeta = \epsilon^{1/2}(x-v_{0}t),~~~~\tau = \epsilon^{3/2}t,
\end{equation}
where $\epsilon$ is a smallness parameter measuring the weakness of the 
perturbation and $V_{0}$ represent the phase velocity of the DAW.
In terms of $\zeta$ and $\tau$ the equations become,
\begin{equation}
 \epsilon^{3/2} \frac{\partial n_{d}}{\partial \tau} - V_{0} \epsilon^{1/2} \frac{\partial n_{d}}{\partial \zeta}+ \epsilon^{1/2} \frac{\partial n_{d} u_{d}}{\partial \zeta} = 0,
  \label{eq:Nor_continuity1}
 \end{equation}
 \begin{eqnarray}
 \begin{split}
\epsilon^{3/2} \frac{\partial u_{d}}{\partial \tau} - V_{0} \epsilon^{1/2} \frac{\partial u_{d}}{\partial \zeta}+ \epsilon^{1/2} u_{d}\frac{\partial u_{d}}{\partial \zeta} = \mu \epsilon^{1/2} \frac{\partial\phi}{\partial \zeta} \\ - \frac{\alpha}{n_{d}} \frac{\partial n_{d}}{\partial \zeta},
 \label{eq:Nor_xmomentum1}
\end{split}
\end{eqnarray}
and 
\begin{eqnarray}
  \begin{split}
  \epsilon^{3/2} \frac{\partial^2 \phi}{\partial \zeta^2} = \frac{1}{\mu} \Bigg\{ \delta_{e} \left[ 1+\sigma_{i} \phi + \frac{1}{2} \sigma^2_{i} \phi^2 + ..\right] \\ - \delta_{i} \left[ 1-\phi + \frac{1}{2} \phi^2 + ..\right] +  n_{d}\Bigg\}.
\label{eq:Nor_possion1}
\end{split}
\end{eqnarray}
  
We can now expand the variables $n_{d}$, $u_{d}$ and $\phi$ in the power series 
of $\epsilon$,
 \begin{eqnarray}
  \begin{split}
n_{d} = 1 + \epsilon n^{(1)}_{d} + \epsilon^{2} n^{(2)}_{d} + ........... \\ 
u_{d} =  \epsilon u^{(1)}_{d} + \epsilon^{2} u^{(2)}_{d} + ........... \\
\phi = \epsilon \phi^{(1)} + \epsilon^{2} \phi^{(2)} + ...........
\label{eq:expansion}
\end{split}
\end{eqnarray}
Substituting Eq.~(\ref{eq:expansion}) into the Eq.~(\ref{eq:Nor_xmomentum1}), 
(\ref{eq:Nor_continuity1}) and (\ref{eq:Nor_possion1}), and equating 
coefficients of $\epsilon^{3/2}$, we get, to lowest order 
\begin{equation}
-V_{0}\frac{\partial n^{(1)}_d}{\partial \zeta} + \frac{\partial u^{(1)}_d}{\partial \zeta} = 0,
\end{equation}
\begin{equation}
-V_{0}\frac{\partial u^{(1)}_d}{\partial \zeta} + \alpha \frac{\partial n^{(1)}_d}{\partial \zeta} - \mu \frac{\partial \phi^{(1)}}{\partial \zeta} = 0, 
\end{equation}
and
\begin{equation}
-\frac{h_{1}}{\mu} \phi - \frac{1}{\mu} n^{(1)}_{d}=0,
\label{eq:h1}
\end{equation}
where $h_{1}$ = $(\sigma_{i} \delta_{e}  + \delta_{i})$ and $\kappa^{2}$  = $\frac{h_{1}}{\mu}$ = $\frac{a^2}{\lambda^2_{D}}$. Consequently the 
Eq.~(\ref{eq:h1}) becomes,
\begin{equation}
- \kappa^2 \phi - \frac{1}{\mu} n^{(1)}_{d}=0,
\end{equation}

After integrating and re-arranging terms, the following linear expressions are obtained,
\begin{equation}
u^{(1)}_{d} = - \frac{\mu V_{0}}{(V^2_{0}-\alpha)} \phi^{(1)},
\label{eq:linear velocity}
\end{equation}
\begin{equation}
n^{(1)}_{d} = - \frac{\mu}{(V^2_{0}-\alpha)} \phi^{(1)},
\label{eq:linear density}
\end{equation}
and
\begin{equation}
V_{0} = \left( \frac{\mu}{h_1} + \alpha \right)^{1/2}.
\label{eq:phase velocity}
\end{equation}
Eq.~(\ref{eq:phase velocity}) represents the phase velocity of the dust 
acoustic wave as a function of $\alpha$. 

Similarly, equating coefficients of $\epsilon^{5/2}$ from 
Eq.~(\ref{eq:Nor_xmomentum1}), (\ref{eq:Nor_continuity1}) and that 
of $\epsilon^{2}$ from Eq.~(\ref{eq:Nor_possion1}), the following equations
are obtained,
\begin{equation}
\frac{\partial n^{(1)}_d}{\partial \tau} -V_{0} \frac{\partial n^{(2)}_d}{\partial \zeta} + \frac{\partial u^{(2)}_d}{\partial \zeta} + \frac{\partial u^{(1)}_d n^{(1)}_d}{\partial \zeta} = 0, 
\label{eq:HO_continuity}
\end{equation}
\begin{eqnarray}
  \begin{split}
\frac{\partial u^{(1)}_d}{\partial \tau} -V_{0} \frac{\partial u^{(2)}_d}{\partial \zeta} + \alpha\frac{\partial n^{(2)}_d}{\partial \zeta} - \alpha n^{(1)}_d\frac{\partial n^{(1)}_d}{\partial \zeta} \\ + u^{(1)}_d \frac{\partial u^{(1)}_d}{\partial \zeta}- \frac{\partial \phi^{(2)}}{\partial \zeta} = 0,
\label{eq:HO_momantum}
\end{split}
\end{eqnarray}
and
\begin{eqnarray}
  \begin{split}
\frac{\partial^{2} \phi}{\partial \zeta^{2}} - \left(\frac{h_{1}}{\mu}\right)\phi^{(2)} + \left(\frac{h_{2}}{2\mu}\right)(\phi^{(1)})^2 - \frac{1}{\mu} n^{(2)}_d= 0,
\label{eq:HO_possion}
\end{split}
\end{eqnarray}
 where $h_{2}$ = ($\delta_{i} -\delta_{e} \sigma_{i}^{2}$). 
Eliminating $u^{(2)}_{d}$, $n^{(2)}_{d}$ and $\phi^{(2)}$ from 
Eq.~(\ref{eq:HO_continuity})-(\ref{eq:HO_possion}) and making use of 
Eq.~(\ref{eq:linear velocity})-(\ref{eq:phase velocity}), we find that 
$\phi^{(1)}$ satisfies the well known KdV equation,
\begin{eqnarray}
  \begin{split}
\frac{\partial\phi^{(1)}}{\partial \tau} + A \phi^{(1)}\frac{\partial\phi^{(1)}}{\partial \zeta} + B \frac{\partial^{3} \phi^{(1)}}{\partial \zeta^{3}} = 0,
\label{eq: KdV equation_SC}
\end{split}
\end{eqnarray}
where the nonlinear coefficient $A$ and the dispersion coefficient $B$ are 
given by,
\begin{eqnarray}
  \begin{split}
A = \left[ \frac{\mu \alpha}{2 V_{0} ( V^2_{0} - \alpha)} - \frac{3 \mu V_{0}}{2 ( V^2_{0} - \alpha)} + \frac{(V^2_{0} -\alpha) h_{2}}{2 V_{0} h_{1}}\right]
\label{eq: A_coeff.}
\end{split}
\end{eqnarray}
\begin{eqnarray}
  \begin{split}
B = \frac{(V^2_{0} - \alpha)\mu}{2 V_{0} h_1}
\label{eq: B_coeff.}
\end{split}
\end{eqnarray}
with $h_{1} = (\sigma_{i} \delta_{e} + \delta_{i})$ and $h_{2} = (\delta_{i} - \sigma^2_{i} \delta_{e} )$.

Eq.~(\ref{eq: KdV equation_SC}) can be solved by separation of variables, 
giving a solution in the laboratory frame as,
\begin{eqnarray}
  \begin{split}
\phi(x,t) = \phi_{m} \text{sech}^{2}\left[\frac{\eta}{\Delta}\right],
\label{eq: Solution_kdv}
\end{split}
\end{eqnarray}
where
\begin{eqnarray}
  \begin{split}
	  \phi_{m} =  \frac{3U_{0}}{A}~~{\rm and}~~\Delta  = \sqrt \frac{4B}{U_{0}},
\label{def-phim-delta}
\end{split}
\end{eqnarray}
are the amplitude and width of the soliton, respectively, $\eta$ is a 
coordinate in the laboratory frame, and $U_{0}$ is the normalized velocity 
of the solitary wave. 
In the limit of $\alpha$ = 0, these coefficients reduce to well know results, A = $\frac{V^3_{0}}{2}\left( h_{2}- \frac{3}{V^4_{0}}\right)$, B = $\frac{V^3_{0}}{2}$, which correspond to the  weak coupling limit 
of the dusty plasma \citep{shukla2015introduction}.
%
%
%
\subsection{Computation of isothermal dust compressibility from QLCA basics \label{compressibility}}
For treating the general high screening regime of the nonlinear solutions,
as central to the present paper, we will be using the more accurate form
(\ref{eq:D_matrix}) of the dynamical matrix $D_{L,T}$. However, as presented 
below, the 
derivation of $D_{L,T}$ in the analytical form is possible in the weak 
screening limit to be compared against the nonlinear results from the GH 
model \cite{sharma2014head} which uses, and is limited to, the weak screening 
limit. 
For this purpose, one might use relationship of $\alpha$, for example, with the systems correlation energy \citep{hou2004,hou2009}.
In the long-wavelength regime ($ka \le 2$) as considered in the present case, 
a model simpler than involving the spectral representations (\ref{eq:D_matrix})
or (\ref{eq:Full_D}) of $D_{L}$ is adopted 
here for $\alpha$ which essentially reproduces the 
more general QLCA results presented further below and therefore is equivalent 
to it \citep{hou2009}. This is realizable 
because $\alpha$ = $\lim_{k \to 0} D_{L}/k^{2}$ \citep{Golden,hou2009}
%
and as long as 
$g(r)$ does not vary significantly with $\kappa$, $D_{L}$ - matrix 
remains suitably expressible in term of correlation energy and its derivative,
given as \citep{rosenberg1997,Golden},
\begin{equation}
	D_{L}({k \to 0}) = \frac{4}{45} \omega^{2}_{pd}\left[1-\kappa \frac{\partial}{\partial \kappa}+ \frac{3}{4} \kappa^2 \frac{\partial}{\partial \kappa^2}\right]\frac{E_{c}(\kappa)}{T_{d} \Gamma} k^2. 
\label{eq:D_k_zero}
\end{equation}
Here $E_{c}$ is the grain-grain correlation energy which can be obtained 
analytically under the long-wavelength 
limit \citep{rosenberg1997,Golden} for a Yukawa fluid, as,
%
\begin{equation}
\frac{E_{c}}{T_{d}}  = a(\kappa) \Gamma + b(\kappa)\Gamma^{1/3} + c(\kappa) + d(\kappa) \Gamma^{-1/3},
\label{eq:E_C}
\end{equation}
where the coefficient up to order $\kappa^{4}$  are given by,
\begin{eqnarray}
\nonumber
a(\kappa)- \frac{\kappa}{2} =  - 0.899-0.103\kappa^2 + 0.003\kappa^4  \\ \nonumber =  a_{0}+a_{2}\kappa^2 + a_{4}\kappa^4, \nonumber
\end{eqnarray}
\begin{eqnarray}
\nonumber
b(\kappa) =   0.565-0.026\kappa^2 - 0.003\kappa^4 = b_{0}+b_{2}\kappa^2 + b_{4}\kappa^4,  \\ \nonumber
c(\kappa) =  - 0.207-0.086\kappa^2 + 0.018\kappa^4 = c_{0}+c_{2}\kappa^2 + c_{4}\kappa^4, \\ \nonumber
d(\kappa) =  - 0.031-0.042\kappa^2 - 0.008\kappa^4 = d_{0}+d_{2}\kappa^2 + d_{4}\kappa^4, \\ 
\end{eqnarray}\label{eq:coefficient}
The expression for the $\alpha$ can be obtained by using the value of these 
coefficients in the equation (\ref{eq:E_C}) and then using 
Eq.~(\ref{eq:D_k_zero}) \citep{rosenberg1997}, 
\begin{eqnarray}
\nonumber
\alpha (\kappa, \Gamma) = \lim_{k \to 0} D_{L}(k)/k^{2} \approx \frac{4}{45 \Gamma}\left[\left(a_{0} + a_{2} \frac{\kappa^2}{2} 
\right.\right.\\ \nonumber
\left.\left.
 + 6a_{4} \kappa^4 \right)\Gamma + \left(b_{0} + b_{2} \frac{\kappa^2}{2} + 6b_{4} \kappa^4\right)\Gamma^{1/3} + \right.\\  
  \left. \left(c_{0} 
  + c_{2} \frac{\kappa^2}{2} \nonumber
   + 6c_{4} \kappa^4\right) + \left(d_{0} + d_{2} \frac{\kappa^2}{2} 
  + 6d_{4} \kappa^4\right)\Gamma^{-1/3}\right].
\end{eqnarray}\label{eq:coefficient}
%
%
%
\begin{figure}[!htbp]
 \centering
 \includegraphics[width=94mm]{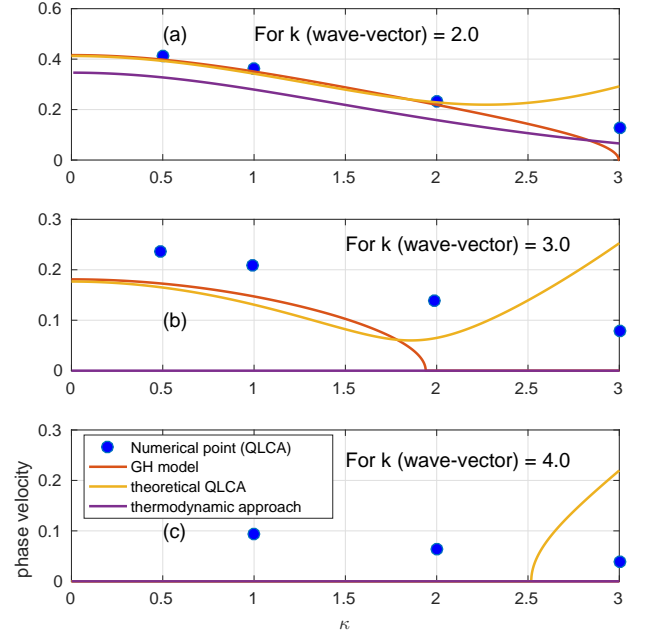}
	\caption{The variation of the linear phase velocity of the DAW  with $\kappa$, (a) for fix  value of $k$ = 2.0, (b) for fix  value of $k$ = 3.0 and (c) for fix  value of $k$ = 4.0.  The red, yellow  and violet curve crossespond to the results of the  GH model, analytical QLCA based model and thermodynamic approach, respectively. The numerical blue points are  plotted using QLCA based model.}
 \label{phase-velocity}
\end{figure} 
One finds, to leading order in $\Gamma$ \citep{Golden},
\begin{equation}
\alpha(\kappa) =  - 0.0799 - 0.0046 \kappa^2 + 0.0016 \kappa^4
\label{eq:alpha}
\end{equation}
%
\begin{figure}[!htbp]
 \centering
 \includegraphics[width=94mm]{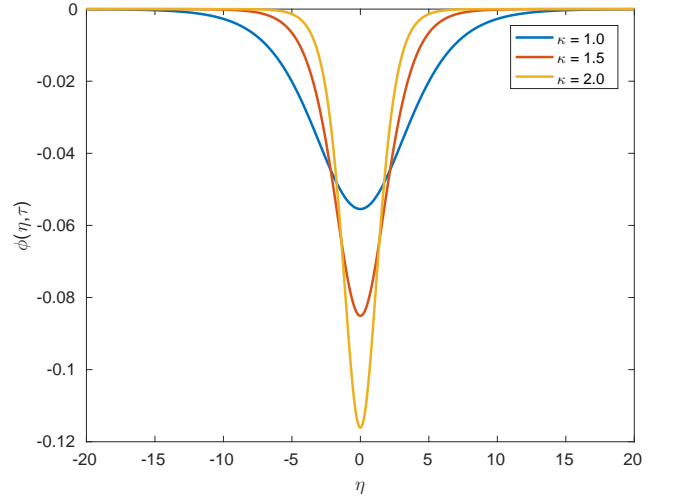}
	\caption{The variation  of the negative solitary potential profile with $\kappa$ in the analytical QLCA limit (equivalent to GH results), for $h_{1}$ = 4, $h_{2}$ = 3.9, $U_{0}$ = 0.1 and inidicated $\kappa$ values.}
 \label{sec}
\end{figure}
%
%
%
%
This dependence of $\alpha$ on parameters $\Gamma$ and $\kappa$ defines the 
effects of dust-dust correlation in a Yukawa system viz. dusty plasma. 
Since $\alpha(\Gamma,\kappa)< 0$, phase velocity of the dust acoustic wave 
in a strongly coupled limit given by Eq.~(\ref{eq:phase velocity}) is 
reduced \citep{rosenberg1997} as presented in  Fig.~\ref{phase-velocity}.
The nonlinear effects in the strongly coupled Yukawa system treated here 
under the QLCA theory can now enter the KdV equation through the coefficients 
$A$ and $B$ which are also the functions of $\alpha(\Gamma,\kappa)$. 
%
%
\section{Numerical pseudo-spectral computation procedure and nonlinear coherent excitations\label{simulation}}
The spatiotemporal domain rather than the 
spectral (Fourier) domain is adopted 
for computationally solve the full nonlinear model described by the set of 
fluid equations 
(\ref{continuity})-(\ref{poisson-equation}),
via a full nonlinear pseudo-spectral approach 
in which the spatial and temporal discretization are possible using the 
independent variables $k$ and $t$, respectively, yet, suitably providing 
the final solutions in the spatiotemporal domain $x$-$t$.
The nonlinear term is calculated by 
implementing the 2/3-truncate rule in order to  avoid the aliasing error \citep{coutsias1989spectral}.
%
%
%
After due verification of the numerical simulation procedure for standard 
nonlinear problems with known analytic solutions (e.g., in $\alpha=0$ limit of the model), 
the localization is included 
via full spectral version of $D_{L}$ (\ref{eq:D_matrix})
and (\ref{eq:Full_D})
to investigate the evolution of the finite amplitude dust acoustic 
KdV-like solitons. 
 
First, the phase velocity of low amplitude sinusoidal perturbations 
evolved by the nonlinear pseudo-spectral computations augmented with 
(\ref{eq:D_matrix}) is compared in Fig.~$\ref{phase-velocity}$ with that 
obtained from the GH model, thermodynamic approach as well as from the 
computations where rather 
approximated (\ref{eq:D_k_zero}) model for $D_{L}$ was used.
%
%
While the general, full QLCA version of $D_{L}(k)$ used in numerical 
pseudospectral simulation (blue points) consistently provides estimates at
higher $k$, the approximate analytical version 
(\ref{eq:D_k_zero}) (effectively underlying both GH and analytical QLCA) 
shows disagreement and, moreover, missing solutions over increasingly 
large range of $\kappa$ values. 
The red and yellow curves 
corresponding to phase velocity in GH and analytical QLCA based model, 
respectively, for the mode $k$ = 2 plotted in 
Fig.~$\ref{phase-velocity}$(a) 
show agreement with phase velocity in the general numerical model 
(\ref{eq:D_matrix}) whereas the thermodynamic approach predicts somewhat 
distinct value for all value of $\kappa$. For mode $k$ = 3 mode plotted in 
Fig.~$\ref{phase-velocity}$(b) the phase velocity from GH, analytical QLCA 
model and thermodynamic approach all begin to show a deviation form the 
Numerical model (QLCA) results. Moreover, the analytical models do not 
predict excitable structures beyond $k$ = 4 for nearly full range of 
$\kappa$, as presented in Fig.~$\ref{phase-velocity}$(c).   

The missing response outlined by Fig.~\ref{phase-velocity} at high $k$ 
values has finite implications for nonlinear solutions which nevertheless
remain obtainable in the form of solutions (\ref{eq: Solution_kdv}) of 
the KdV equation \cite{sharma2014head}, despite a bulk of high $k$ 
constituents not contributing to their construction owing to the 
limitations of the underlying approximation.
Consequently, the evolution of nonlinear structures indeed shows different 
physical characteristics and considerably sensitivity to 
$\kappa$ variation only when obtained by means of the present, more general, 
numerical QLCA implementation (used for blue dots in 
Fig.~$\ref{phase-velocity}$) as analyzed below using coherent perturbations. 

\begin{figure}[!htbp]
 \centering
 \includegraphics[width=96mm]{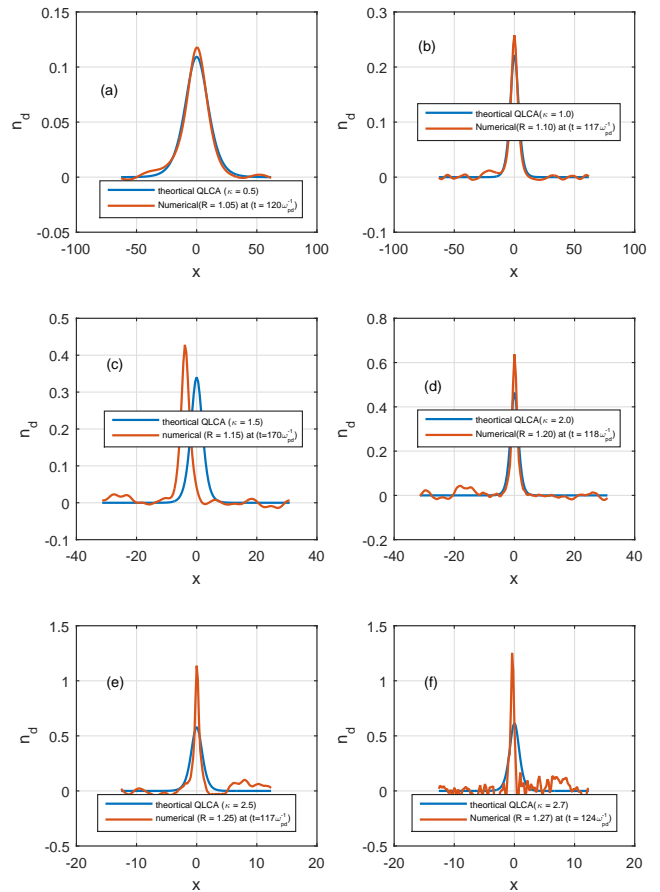}
\caption{ The solitary density profile with, (a) $\kappa$ = 0.5, (b) $\kappa$ = 1.0, (c) $\kappa$ = 1.5, (d) $\kappa$ = 2.0, (e) $\kappa$ = 2.5, (f) $\kappa$ = 2.7,  in both the numerical and analytical limit of the QLCA theory, for $h_{1}$ = 4, $h_{2}$ = 3.9, $U_{0}$ =0.1  }
 \label{sech_com}
\end{figure}

As presented in Fig.~\ref{sech_com}, 
the general present pseudospectral simulations (red) initialized using 
the solutions (\ref{eq: Solution_kdv}) of the KdV equation
as initial condition (stationary blue profiles) show an initial transient 
until the solitary structures settle down to newer stably propagating 
solitary profiles which are distinct from what prescribed by the analytical 
approximation based high $\kappa$ QLCA solutions (which agree with GH 
solutions). 
These newer coherent solitary structures are far steeper than those predicted 
by the analytical approximation underlying the KdV equation.
At relatively higher $\kappa$ values the numerical solution tend to grow 
extreamly steep and acquire cusp-like spatial profiles showing weaker 
temporal coherence.
The accessibility to their evolution with enough accuracy is thus 
seen to also depend on the numerical resolution of the computations.
%
%
Remarkably, this initial transient is not produced in preseudospectral
numerical evolution in the low $\kappa$, or weak screening, regime and the
initial profiles prescribed by the analytical KdV solutions propagate
without any significant distortion, indicating that both analytical QLCA 
and the GH model make a reasonably good prediction of the nonlinear coherent 
structures, remaining in agreement with the general numerical 
implementation.
\begin{figure}[!htbp]
 \centering
\includegraphics[width=90mm]{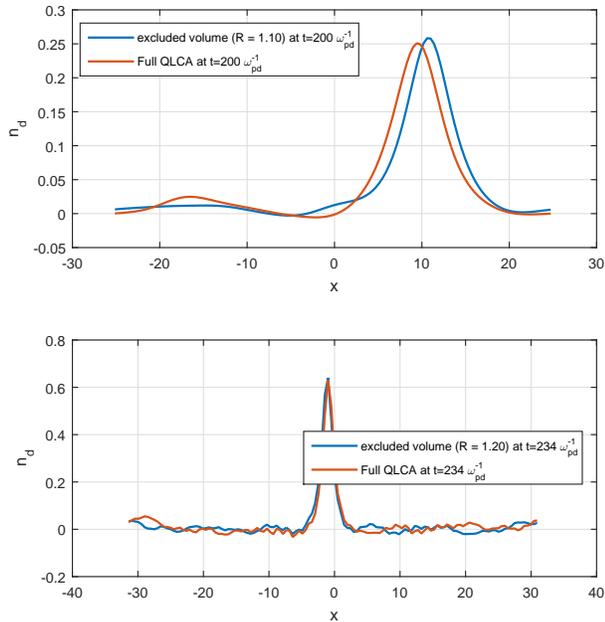}
 \caption{ The solitary density profile with, (a)  Excluded volume (R) = 1.10, (b) Excluded volume R = 1.20, in both the Excluded  volume approximation  and Full QLCA  $D_{L}$ matrix, for $h_{1}$ = 4, $h_{2}$ = 3.9, $U_{0}$ = 0.1.  }
 \label{R_FULL}
\end{figure}
Additionally, the numerical approaches using $D_{L}$ based on 
(\ref{eq:D_matrix}) and (\ref{eq:Full_D}) were also compared and 
found to be nearly in agreement with each other.
The red and blue curve, in Fig.~\ref{R_FULL} (a), correspond to the solition 
structure obtained from the full QLCA $D_{L}$ matrix with $\kappa$ = 1.0 and 
excluded volume approximation with R = 1.10, respectively. There is no 
significance difference observed to be developed, accordingly, in the solition 
structure. 
Similar behavior of the solition structures can also been seen from  
Fig.~\ref{R_FULL}(b) at relative higher screening parameter value 
$\kappa$ = 2.0 in which the solition structures obtained from both the full QLCA $D_{L}$ matrix with $\kappa$ = 2.0 and Excluded volume approximation with R = 1.20, respectively, almost overlap.

The agreement between GH model solutions with the numerical computations 
in smaller $\kappa$ regime relates, more quantitatively, to the 
characteristics of linear modes for lower and higher $k$ values, which is  
analyzed over a range of $\kappa$ values, for example in 
Fig.~\ref{phase-velocity}. Only a moderate variation of 
$g(k)$ with $\kappa$ for lower $k$ means that $D(k \rightarrow 0)$ can 
still be derived from the system correlation energy and its derivatives. 
No analytical approximation is however possible for relatively 
higher values of $k$ and $\kappa$, and $D_{L}$ remains no longer 
correctly expressible in terms of the system correlation energy. 
On the other hand, the general adoptations of $D_{L}$, either 
(\ref{eq:D_matrix}) or (\ref{eq:Full_D}), by means of our pseudospectral 
(spatiotemporal, yet fully resolved in the $k$ values of modes constituting 
the nonlinear structures) numerical model, accounting for the structural 
effects continues to describe the smaller wavelengths and 
relative stronger screened excitations in this limit.
%
%
%
Therefore, Numerical simulation accounting for the full structural effects can suitably describe the higher wavelengths and relative stronger screened excitations in such systems.

%
%
%
%
\section{Characterization of analytical nonlinear solitary solutions in weak screening limit\label{results}}
In order to analyze solitary waves in a strongly coupled but weakly screened dusty 
plasma with finite localization in rather detail, we begin by presenting the amplitude 
$\phi_{m}$ and 
width $\Delta$ of the soliton given by Eq.~(\ref{def-phim-delta}) 
(besides normalized velocity of the $U_{0}$). The amplitude $\phi_{m}$ and 
width $\Delta$ depend on $\alpha$
and, in turn, on $\kappa$ and $\Gamma$ through factors $A$ and $B$ given by 
Eq.~(\ref{eq: A_coeff.}) and (\ref{eq: B_coeff.}), respectively.
%
 The negative solitary potential profile  is plotted in Fig.~\ref{sec} 
in the strong coupling (QLCA) limit with different values of screening parameter $\kappa$. For the constant value of $U_{0}$,
\begin{figure}[!htbp]
 \centering
 \includegraphics[width=94mm]{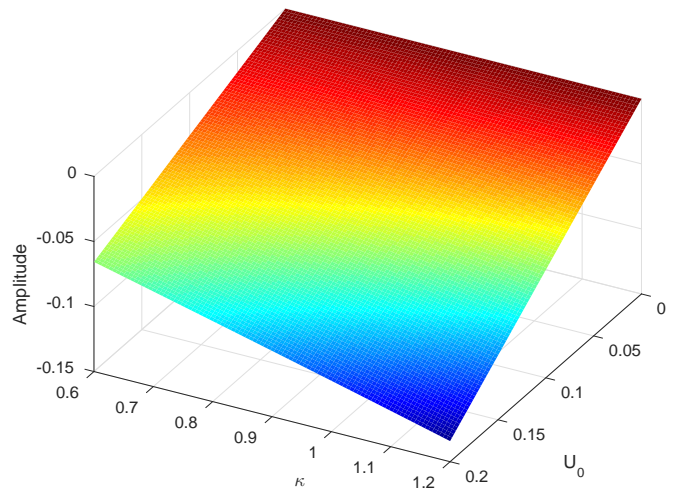}
  \caption{The variation of the amplitude of the negative solitary potential with parameters $\kappa$ and $U_{0}$ for constant value of $h_{1}$ = 4, $h_{2}$ = 3.9 .}
 \label{amplitude_2d}
\end{figure}
\begin{figure}[!htbp]
 \centering
 \includegraphics[width=94mm]{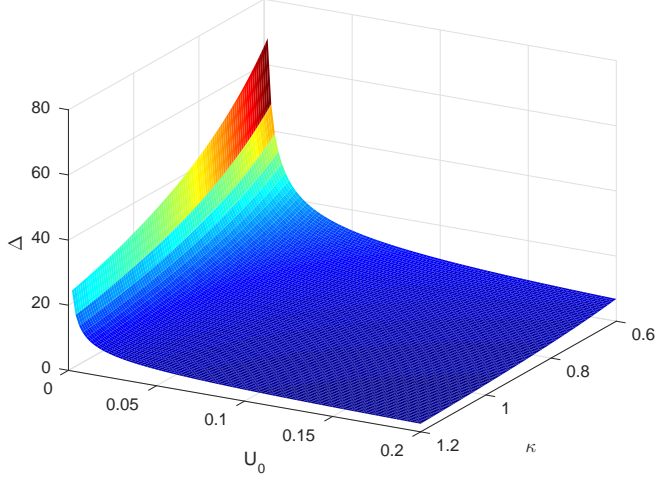}
  \caption{The variation of the width of the negative solitary potential with parameters $\kappa$ and $U_{0}$ for constant value of $h_{1}$ = 4, $h_{2}$ = 3.9.}
 \label{width_2d}
\end{figure}
the amplitude of the solitary wave increases with $\kappa$ in
strongly coupled (QLCA) limit since the amplitude has a inverse dependence 
on with factor $A$ reduces with $\kappa$ in this limit. 
The width being directly proportional to factor $B$ reduces with $\kappa$ in 
 the strongly coupled  (QLCA) limit. 
 %
This behavior of solitary waves explored for the strongly coupled and 
localized phase of the dust (adopting QLCA based approach) is in contrast to 
the effects of strong coupling on a dust solitary wave analyzed in a model 
where strong-coupling effects are introduced via an effective temperature 
$T^{\rm (eff)}$ \cite{cousens2012}. A better agreement the soliton solutions 
recovered in the present treatment is however seen with the Molecular 
Dynamical (MD) simulation 
results \citep{tiwari2015molecular,kumar2017observation}. Moreover, the 
behavior of solitary waves explored by adopting QLCA based approach is also 
in contrast to the effects of dust temperature on a dust solitary wave 
analyzed from the theoretical models 
\citep{mamun2002electrostatic,mamun2005nonlinear} in which the amplitude 
and width of dust acoustic solitary waves decreases and increases, 
respectively, with increasing the dust temperature.

The 2D surface plot of the amplitude $\phi_{m}$ and width $\Delta$ of the 
dust solitary waves are presented in Fig.~\ref{amplitude_2d} and 
\ref{width_2d}, respectively, describing their variation with respect to
variation of the parameters $U_{0}$ and $\alpha$. For a fix value of $\alpha$, 
while the amplitude of the dust solitary wave increases with $U_{0}$ its 
width is shown to decrease, which is a behavior consistent with the well known 
characteristics of the solitons where the product $|\phi_{m}|\Delta^{2}$ is 
independent of $U_{0}$. 
\section{Simulations with initial profiles of general form}
For simulating more general but initially localized nonlinear perturbations, 
we have also launched a more general, Gaussian-shaped, initial density 
perturbations in the pseudospectral simulation procedure given by,
 \begin{figure}[hbt]
 \centering
 \includegraphics[width=94mm]{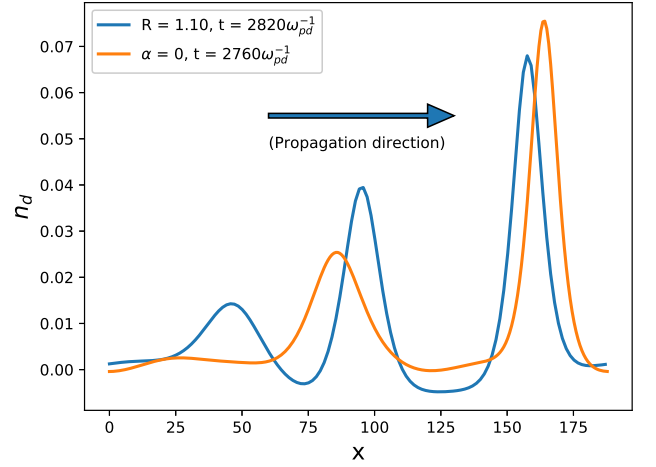}
  \caption{ Time evolution profile of an initial Gaussian dust density perturbation in the both weak and strong coupling limit for $\kappa$ = 1.0, $n_{0}$ = 0.05 and $\Delta$ = 20.}
 \label{gaussian_0p05}
\end{figure}
 \begin{equation}
n_{d} = n_{0} \exp\left[-\left(\frac{x-x_{0}}{\Delta}\right)^2\right], 
\end{equation}
%
Fig.~\ref{gaussian_0p05} presents the evolution of the initial Gaussian density 
perturbation, with $n_{0}$ = 0.05, $\Delta$ = 20 and $x_{0}$ = 64, in both 
strong ($\alpha \ne 0$, $R=1.1$) and weak 
($\alpha=0$) coupling limit of the model, 
It is evident from the presented evolution that in the weakly coupled limit 
the perturbation decays into two unequal solitons, whereas it splits in to
three unequal amplitude soliton structures when the dust-dust correlations are 
accommodated in the model with complete localization effects. It can be seen 
that the  taller soliton propagates with a velocity faster than the smaller 
soliton which remains a well know property of the soliton/solitary wave 
structures of KdV type.  
\begin{figure}[!htbp]
 \centering
 \includegraphics[width=94mm]{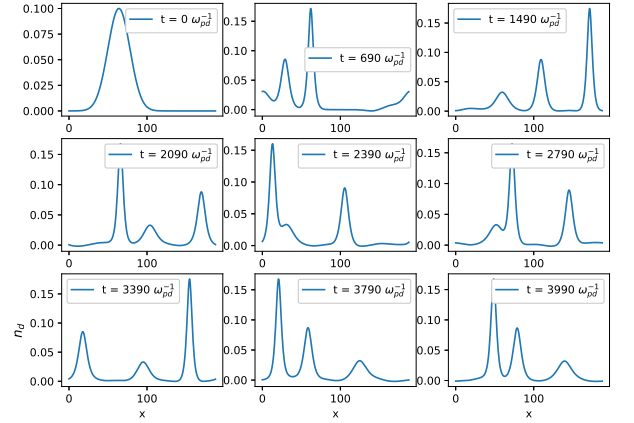}
  \caption{Time evolution of an initial Gaussian dust density perturbation profile in a weakly coupled limit for $\kappa$ = 1.0, $n_{0}$ = 0.1 and $\Delta$ = 20.}
 \label{Gaussian_0p1_weak}
\end{figure}
\begin{figure}[!htbp]
 \centering
 \includegraphics[width=94mm]{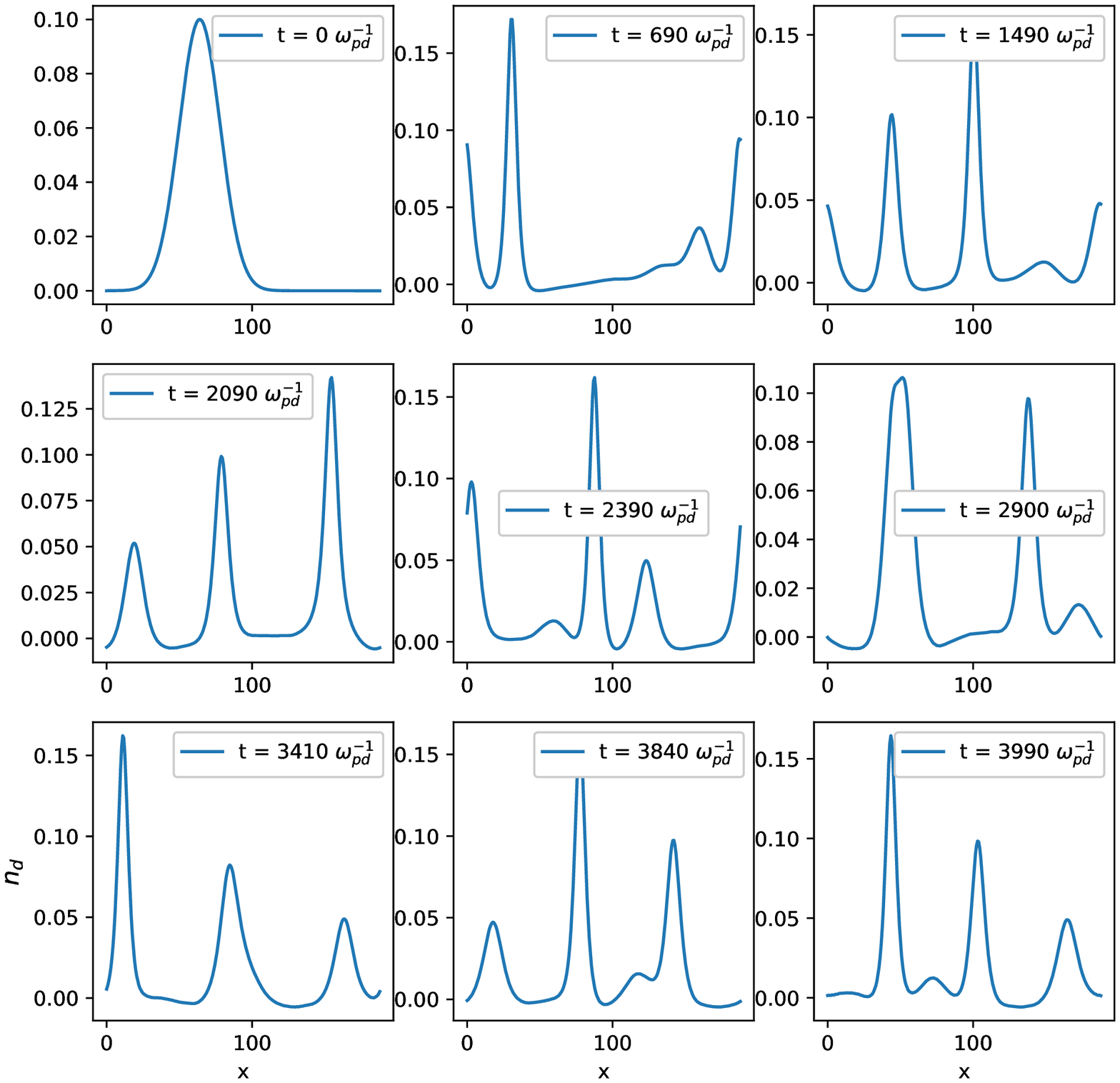}
  \caption{ Time evolution of an initial Gaussian dust density perturbation profile in the strongly coupled limit for $\kappa$ = 1.0, R($\kappa$) = 1.10, $n_{0}$ = 0.1 and $\Delta$ = 20. }
 \label{Gaussian_0p1_strong}
\end{figure}
%
%
\begin{figure}[!htbp]
 \centering
 \includegraphics[width=94mm]{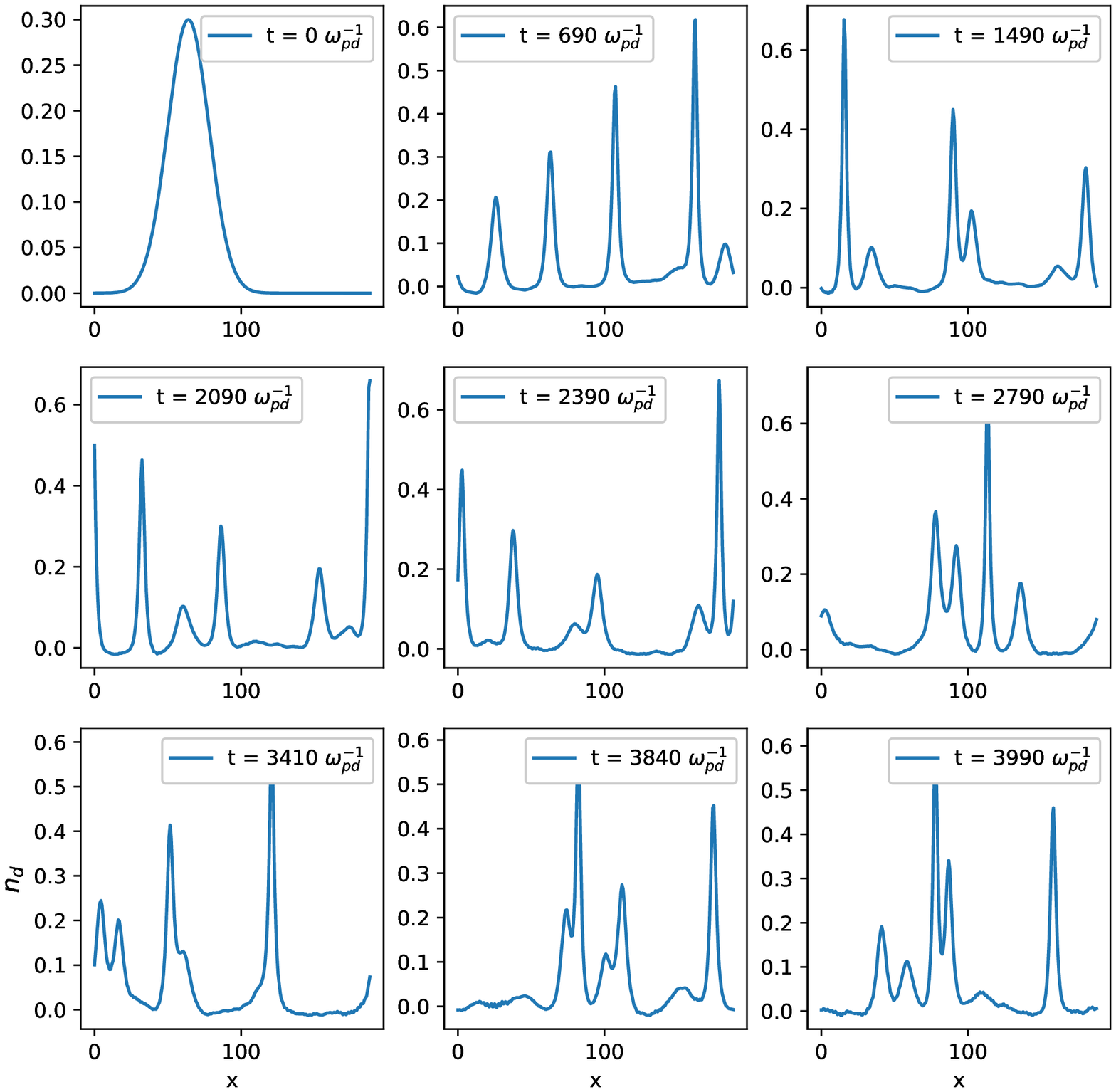}
  \caption{ Time evolution of an initial Gaussian dust density perturbation profile in the strongly coupled limit for $\kappa$ = 1.0, R($\kappa$) = 1.10, $n_{0}$ = 0.3 and $\Delta$ = 20.}
 \label{Gaussian_0p3_strong}
\end{figure} 
 \begin{figure}[!htbp]
 \centering
 \includegraphics[width=94mm]{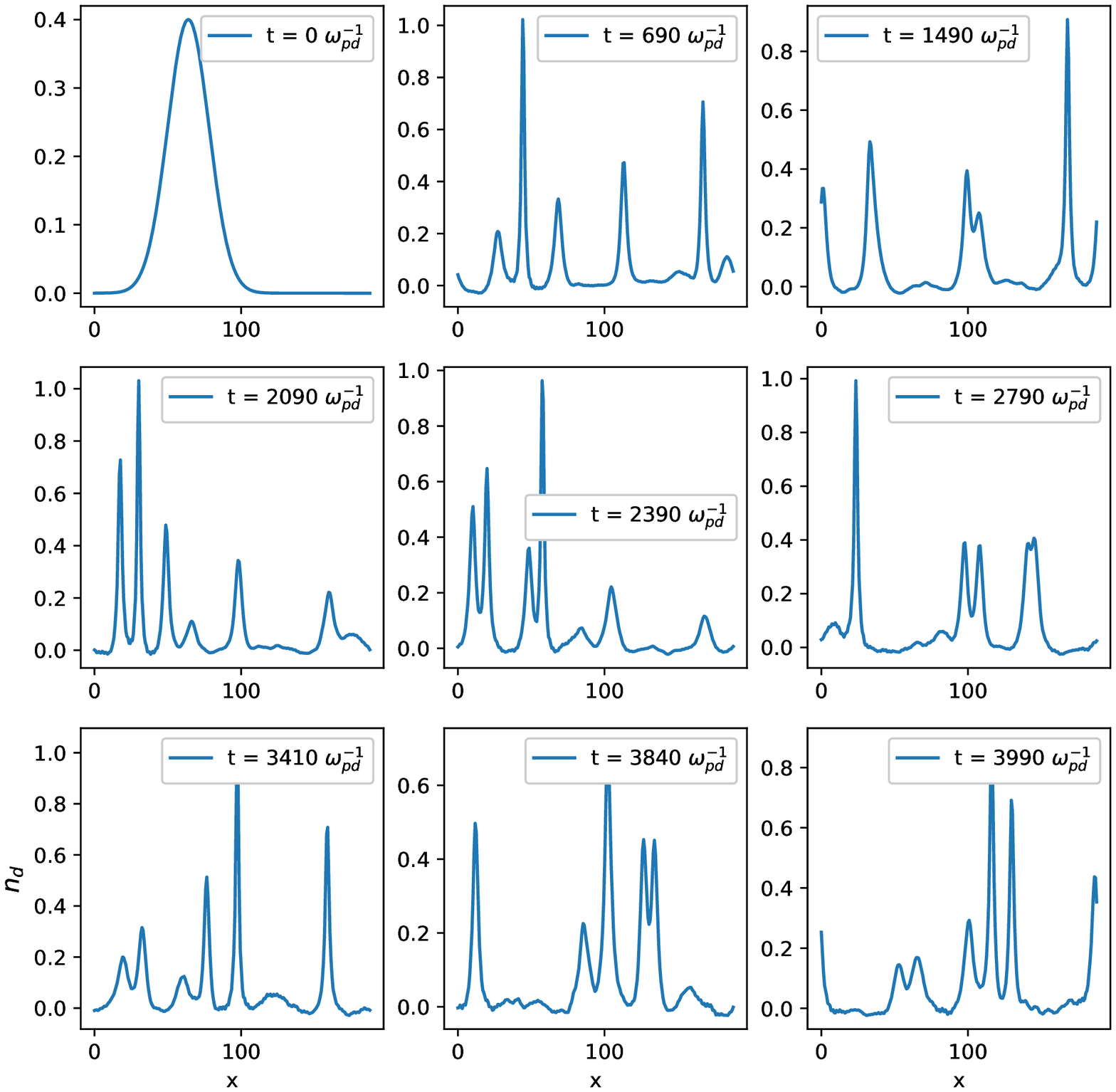}
  \caption{Time evolution of an initial Gaussian dust density perturbation profile in the strongly coupled limit for $\kappa$ = 1.0, R($\kappa$) = 1.10, $n_{0}$ = 0.4 and $\Delta$ = 20. }
 \label{Gaussian_0p4_strong}
\end{figure}
%
%
\begin{figure}[!htbp]
 \centering
 \includegraphics[width=94mm]{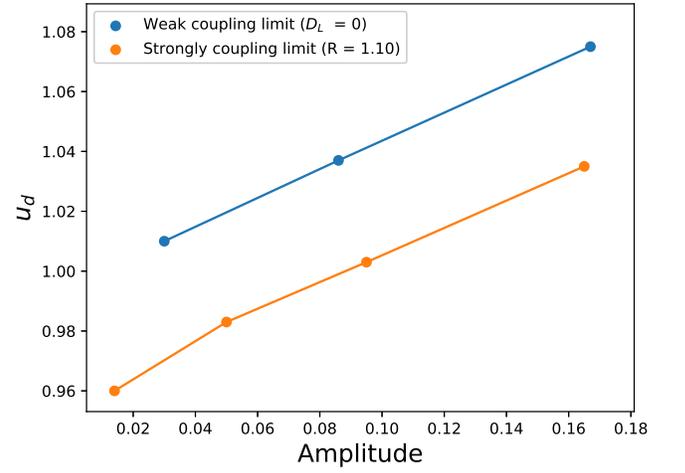}
  \caption{ Mach Number ($u_{d}$) of the dust acoustic solitons as a function of their amplitude in the weak and strongly coupling limit for $\kappa$ = 1.0 and $n_{0}$ = 0.1. }
 \label{velocity}
\end{figure} 
A number of more general initial condition structures with a range of initial 
amplitude are studies further in order to understand the influence, on these 
excitations, of various initial density perturbation profiles. 
Fig.~\ref{Gaussian_0p1_weak} presents the evolution of a density perturbation 
with a relatively higher initial amplitude $n_{0}$ = 0.1 within the 
weak-coupling limit ($\alpha = 0$). 
The emergence of three unequal height soliton has once again been noted in 
this case which appears identical to the strong-coupling case presented in 
Fig.~\ref{gaussian_0p05}. 
It can therefore be concluded that the effect of strong-coupling is 
somewhat equivalent to the choice of an increased initial amplitude of the 
perturbation. The next Fig.~\ref{Gaussian_0p1_strong} presents the evolution 
of the density perturbation with the initial amplitude $n_{0} = 0.1$ however 
this time in the strong-coupling limit. What is noted is the emergence of 
a train of four solitons of orderly reducing heights, 
confirming the equivalence of 
strong-coupling to the initial amplitude of the perturbation. 
Since these solitary structures  visibly preserve their identities for a 
considerably longer time (i.e., several orders of inverse dust acoustic 
frequency) and even after undergoing mutual interaction (or collisions), 
they indeed represent the dust acoustic soliton structures. 
Fig.~\ref{Gaussian_0p3_strong} 
and \ref{Gaussian_0p4_strong} demonstrate the reasonably consistent solitary 
evolution of the initial density perturbation with even higher amplitudes, 
$n_{0} = 0.3$ and $n_{0} = 0.4$, respectively, up to a sufficiently longer 
evolution time $\omega_{pd} t = 3900$.
\begin{figure}[!htbp]
 \centering
 \includegraphics[width=94mm]{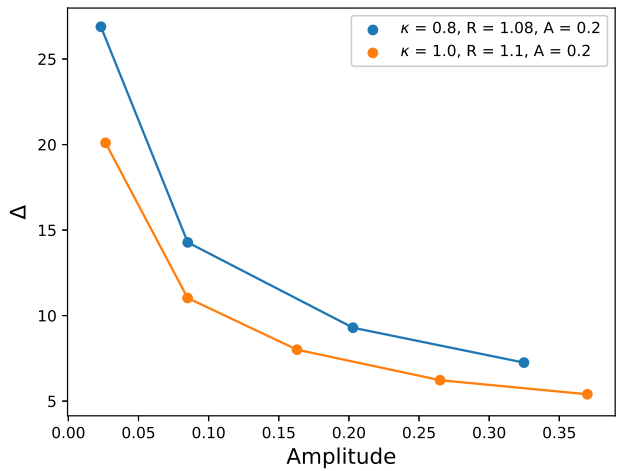}
  \caption{Width ($\Delta$) of the dust acoustic solitons as a function of their amplitude in the strong coupling limit for different value of parameter R($\kappa$).}
 \label{width}
\end{figure}

In order to make a comparison with the theoretical estimates presented in 
Sec.~\ref{qlca}, the normalized velocity (to linear dust acoustic velocity) 
and width of the dust acoustic solitons are presented in 
Figs.~\ref{velocity} and \ref{width}, respectively. These are measured for 
different excitation amplitudes after the individual solitons appear 
in both weak and the strong coupling cases. The Mach number  $u_{d}$ of the solitary wave increases linearly with
amplitude in both strong and weakly coupled limit consistent with the experimental observations\cite{sharma2014head,nosenko2002nonlinear,bandyopadhyay2008experimental}.
The measured Mach number $u_{d}$ 
of the solitons in the strong-coupling is smaller than in the weak coupling 
($\alpha$ = 0) limit of the model. The effect of the strong dust-dust 
correlation in reducing the phase velocity of the solitons, is consistent 
with the theoretical predictions of Sec.~\ref{qlca}. For a fixed value of 
$R(\Gamma$,$\kappa)$ (the excluded volume parameter) the width of a soliton 
decreases with increasing the amplitude, as can be seen from  
Fig.~\ref{width}. On the other hand, for a fixed value of the soliton 
amplitude, the width of a soliton decreases with increasing the 
dust-dust correlation via parameter $\kappa$ or $R(\Gamma$,$\kappa)$. 
This effect of the dust-dust correlation is also in correspondence with the 
above theoretical prediction of Sec.~\ref{qlca}.  
\begin{table}[h!]
\centering
\begin{tabular}{||c c c ||} 
 \hline
 Amplitude(A) & Widht($\Delta$) & (A$\Delta^{2}$) \\ [0.5ex] 
 \hline\hline
 0.0235 & 26.90 & 17.040  \\ 
 0.0851 & 14.28 & 17.350  \\
 0.2029 & 9.292 & 17.518  \\
 0.3248 & 7.251 & 17.077  \\ [1ex] 
 \hline
\end{tabular}
\caption{ The soliton characteristic property for the parameters $\kappa$ = 0.8, R($\kappa$) = 1.08 and A = 0.2.}
\label{table:1}
\end{table}
Finally, in order to present a more quantitative aspect of the presented 
numerical soliton solutions we conclude by presenting another well known 
property of the Kdv solitons, namely, the product of amplitude and squared 
width, $\phi_{m}\Delta^{2}$, of the solitons which remains independent to 
the dust acoustic soliton mach number. Our solutions confirm that this 
property of the dust acoustic soliton is preserved and persists even in 
the realm of the QLCA approach applied to the soliton formulation, as 
presented in the form of data in Table~\ref{table:1}. 
\section{Summary and conclusions \label{conclusion}}
To summarize, the nonlinear approach is made to finite amplitude 
excitations of strongly-coupled fluids such that dust 
localization effects are present and can not be neglected especially while 
treating strong screening of localized charges generated by the background 
plasma, for example in a laboratory dusty plasma. 
The nonlinear excitations treated under analytical approximations that 
exclude localization nevertheless prescribe localized solutions with 
very limited reproducibility by a numerical implementation of more 
full nonlinear model accounting for the localization by means of the 
dynamical matrix $D_{L}$ under the quasi-localized charge approximation.
Inclusion of the finite dust localization, essentially requiring to 
account for the structured pair-correlation function $g(r)$, is suitably 
facilitated by the quasi-localized charge approximation 
framework. With its present applications remaining limited to spectral domain 
and hence largely to exploring the linear dispersions,
the persisting challenge of inclusion of these effects in a full nonlinear 
and therefore spatiotemporal formulation is addressed in the present study. 
This is done by developing and solving an analytical nonlinear model for 
the localized-phase of a strongly coupled dusty plasma system while 
incorporating the elements of detailed QLCA formulation. The nonlinear 
solutions are obtained both by analytical and numerical implementations, 
in a nonlinear pseudspectral approach, of the dynamical matrix $D_{L}$
duly accounting for a structured pair-correlation function $g(r)$. 
The characterization of both periodic and coherent solitary nonlinear 
structures has allowed to identify, in strong screening limit, the 
contribution over a considerable spectral range spectral which is adequately 
accounted for by a more general QLCA implementation of the localization
by means of nonlinear pseudospectral procedure.

Among main results, we have first shown that an analytical 
approximation of the QLCA dynamic matrix 
$D_{L}$\citep{Golden,rosenberg1997}
in terms of excess energy (the OCP implementation), limited to weak screening 
limit, reproduces results available from the GH prescription. 
However, application of the QLCA formulation in its full capacity to the 
strong screening limit subsequently allows recovery of a strong departure 
in the evolution of the nonlinear excitations from that of their weak 
screening counterparts, accessed by OCP limit implementation of the 
procedure well within QLCA framework. Accordingly, the nonlinear 
pseudospectral procedure
initialized with the analytical coherent solitary solutions of the KdV 
equation although propagate unmodified in the small $\kappa$ limit, they
undergo an initial transient and self-consistently settle down to newer 
more steeper profiles for large $\kappa$ values. 
At relatively higher $\kappa$ values the numerical solutions tend to grow 
extremely steep and acquire cusp-like spatial profiles showing weaker 
temporal coherence indicating that they are rather governed by a modified 
KdV equation, as supported by certain recent arguments\citep{hunter1991dynamics,kumar2012longitudinal}.
The observations that the accessibility to their evolution with enough 
accuracy is only limited by the numerical resolution of the computations
indicate potential applicability of the analysis procedure to rather yet 
unexplored limits of the parameter space.

In terms of linear structures, the access to larger frequency, or shorter 
wavelength limit, where the OCP 
based descriptions, including the original GH dispersion as well as its QLCA
counterpart, show strong limitation is now available by means of the presented 
excluded volume approximation of the QLCA dynamical matrix. The corresponding 
QLCA linear dispersion is shown to closly agree with the results of the MD 
simulation in this regime.
In essential quantitative terms, the $\kappa\sim 1$ limit for the existing
Yukawa fluid models of both periodic linear and coherent nonlinear excitations 
to show agreement with the experiment and simulations has been overcome by 
the present QLCA prescription and agreement is now recoverable up to 
$\kappa\sim 2.7$. 

Among its major limitations, the QLCA approach remains
unsuitable to an arbitrarily large $\kappa$. It also loses its
applicability when approaching the hard sphere interaction limit 
\citep{khrapak2017collective}.          
The nonlinear processes involving unstable density fluctuations of shorter 
wavelength, for example, DAW suffer a modulational instability at short 
wavelengths, can however be captured by the QLCA model and a related nonlinear
analysis is being communicated by the authors separately, which also forms a 
suitable future work to the present study.
\section{ACKNOWLEDGMENT}
The simulation work presented here is
performed on ANTYA cluster at the Institute for Plasma Research (IPR), Gandhinagar, India.

\appendix*
\label{A}
\section{Incorporation of the strong coupling effects without localization \label{strong-coupling}}
\subsection{Effective strong coupling in random phase limit}
While explicit representation of strong coupling by retaining the correlation
via dynamical matrix $D_{L,T}$ is central to QLCA approach, a strongly coupled
yet random phased state ($g(r)\rightarrow 1$) is treatable
by an effective representation of the
dust-dust correlation by supplementing the kinetic dust pressure, 
$P_{dk}$, by an 
effective pressure, $P_{di}$, or an isothermal dust compressibility 
~\cite{cousens2012},
%
 \begin{eqnarray}
  \begin{split}
	 \frac{\partial{u}_{{\rm d}x} }{\partial t} + ({\bf u}_{{\rm d}}
	 \cdot\nabla) {u}_{{\rm d}x} 
	 = \frac{q_{\rm d}}{m_{\rm d}}{E}_{x} - \frac{1}{m_{d} n_{d}}\left[\frac{\partial P_{dk}}{\partial x} +\frac{\partial P_{di}}{\partial x}\right]. \\ 
	  \label{momentum-balance-rpa}  
\end{split}
\end{eqnarray}
%
An effective temperature $T_{d}^{\rm (eff)}$, apart from dust kinetic 
temperature is then used to express the copressibility arising from the correlations, 
\begin{eqnarray}
  \begin{split}
	 \frac{\partial{u}_{{\rm d}x} }{\partial t} 
	 = \frac{q_{\rm d}}{m_{\rm d}}{E}_{x} - \frac{T_{d}^{\rm (eff)}}{m_{d} n_{d}}\frac{\partial n_{d}}{\partial x} ,\\ \label{momentum-balance-rpa-sc}
\end{split}
\end{eqnarray}
where,
\begin{eqnarray}
	T_{d}^{\rm (eff)} = \frac{N_{nn}}{3} \Gamma T_{d} (1+\kappa) \exp(-\kappa), 
\end{eqnarray}
which is a few orders of magnitude higher
than the kinetic temperature of the dust. 

Eq.~(\ref{momentum-balance-rpa-sc})
along with the continuity and Poisson equations produces the linear 
dispersion relation of the strongly coupled dusty plasma in the 
random phase limit,
\begin{equation}
\omega^2 = 
	 \left(\frac{d_{1} k^{2} \omega^2_{\rm pd}}{1
	+{k^{2}}}
	\right) + \beta k^2,
	\label{dispersion-rpa}
\end{equation}
where $\beta$ and $d_{1}$ are the model parameters defined by the dust and 
background plasma \cite{cousens2012}. 
\subsection{Thermodynamic approach to the strong-coupling in random phase limit}
 One of the simplest approachs to account for the strong coupling effects in this limit has been presented by Khrapak \cite{khrapak2015fluid} in which strong coupling effects are included in the conventional fluid model by supplementing it with the appropriate thermodynamic functions,
\begin{equation}
c^2_{s} = \frac{\omega^2}{k^2} = 
	 \omega^{2}_{pd}\left(\frac{1}{\kappa^{2}}
	+ \frac{\gamma \mu}{3 \Gamma}\right), 
	\label{dispersion-thermo}
\end{equation}
where $\gamma$ and $\mu$ is the adiabatic index and isothermal compresssibility modulus, respectively. To evaluate the effects of strong coupling on the dispersion relation and sound velocity of Yukawa fluid, the parameter $\gamma$ and  $\mu$ are represented in term of thermodynamic functions of Yukawa fluid as,
 \begin{equation}
\gamma\mu = \mu + \frac{\left[p-\Gamma(\partial p/\partial \Gamma)\right]^{2}}{u-\Gamma(\partial p/\partial \Gamma)},
\end{equation}
where u, p and $\mu$ are the internal energy, pressure and isothermal compresssibility modulus, respectively, which account for the both particle-particle correlation and plasma-related effects \citep{khrapak2015thermodynamics}. These linear dispersion are compared with the more general linear dispersion generated by  QLCA of linear perturbation equation (\ref{linear-equation}), obtained as below. 
With the isothermal compressiblity having negative sign, at  sufficiently strong coupling this results in the loss of solution at large $k$ values, leading to null frequency observed in dispersion relation plotted with a green line in Fig.~\ref{dispersion-comparison}. This issue is effectively resolved by inclusion of localization in QLCA model as discussed further below.  
%

\subsection{Generalized  hydrodynamic model (GH)}
The GH model, based on phenomenological dispersion \citep{kaw2001collective}, incorporates the strongly coupling effects via  nonlocal visco-elasticity with memory effects arising from the strong correlation among constituent particles. In a long wavelength limit, the dispersion relation of strongly coupled yukawa system in their kinetic regime can be written as, 
\begin{equation}
\omega^2(k) = 
	\frac{\omega^2_{\rm pd} k^{\rm 2}}{k^{\rm 2}
	+\kappa^{\rm 2}} + \frac{\omega^2_{\rm pd} k^2}{\Gamma}\left(\frac{1}{3} + \frac{4}{45} u_{ex}\right)
	\label{disperion_GH},
\end{equation}
where $u_{ex}$ is the normalized excess energy given in Ref. \citep{kaw2001collective}.

\bibliographystyle{apsrev4-1}
\bibliography{paper}
\end{document}